\documentstyle[twocolumn,prl,aps]{revtex}

\def\beqn{\begin{eqnarray}}
\def\beqns{\begin{eqnarray*}}
\def\eeqns{\end{eqnarray*}}
\def\beq{\begin{equation}}
\def\eeq{\end{equation}}
\def\bea{\begin{array}}
\def\ea{\end{array}}

\def\<{\langle}
\def\>{\rangle}

\begin{document}

\twocolumn[\hsize\textwidth\columnwidth\hsize\csname @twocolumnfalse\endcsname
\draft
\preprint{}
\title{\sc Fermi surfaces and anomalous transport in quasicrystals}

\author{\sc S. Roche and T. Fujiwara}

\address{Department of Applied Physics, University of Tokyo, 7-3-1 Hongo,
Bunkyo-ku, Tokyo 113, Japan.}

\maketitle

\begin{abstract}
\leftskip 54.8pt
\rightskip 54.8pt
Electronic transport
properties of quasicrystals is discussed theoretically. 
By means of ab-initio Linear Muffin Tin Orbitals ({\cal LMTO})
 calculations, electronic bandstructure and
corresponding Fermi surfaces of several quasicrystalline approximants
are obtained. A criterion for distinguishing between metallic and
anomalous transport properties in intermetallics is proposed.
Unconventional temperature dependence of conductivity of
quasicrystals and approximants is addressed in a second part.
It is shown that power law
exponents can be directly deduced from scaling analysis of the Kubo
formula. Finally in relation to our results,
 we briefly summarize actual knowledge on low temperature transport regimes.
\end{abstract}

\vspace{20pt}

\noindent

\pacs{PACS numbers: 72.90.+y 61.44.Br 72.10.-d }

]

\section{Introduction}

\vspace{20pt}

\hspace{\parindent}In 1984,
 the five-fold symmetric quasicrystals (QC) were discovered
in AlMn based intermetallic alloys.~\cite{AlMn} So far,
a lot of theoretical and experimental
work have been done in order to describe their crystallographic
properties. Concerning physical properties,
electronic and magnetic behaviours have unveiled very interesting
phenomena and somehow disconcerting physics.~\cite{Avignon,Roche-I}
 In particular, contrary to conventional
metallic alloys, the thermal and electrical conductivities of
quasicrystalline phases have been found to be unusually low,
almost as low as those of insulators. Features like strong hardness, low
friction coefficients, resistance against oxydation and corrosion, and optical
properties have turned QC into interesting alternatives for several
technological applications.~\cite{Applications}

\vspace{10pt}

\hspace{\parindent}Theoretical works have been concerned with the
 relation between quasicrystalline order (short, mesoscopic and
long-range) and electronic localization and diffusion 
modes.~\cite{Kohmoto-QP,Macia,Roche-IV,Roche-II} The
electronic structure of real quasicrystals can be deduced from
ab-initio LMTO calculations, performed on realistic
structural models of their approximants.~\cite{Fujiwara-II,Fujiwara-III} A
pseudo-gap at Fermi level and a dense concentration of low dispersive
bands in its close vicinity, constitute the main features of their
 electronic structure.

\vspace{10pt}

\hspace{\parindent}In this paper, the topic of Fermi surfaces of
quasicrystals is addressed. Such a concept may appear meaningless
in quasiperiodic systems for which no reciprocal space can be properly
defined, due to the lack of translational symmetry. Notwithstanding,
quasicrystals share similar physical properties with certain crystalline
alloys called quasicrystalline approximants. Indeed, 
these alloys do approximate quasicrystalline order on
a length scale which actually defines the characteristic size of their
unit cell. Such a family of increasing length approximants
(1/1,2/1,3/2,5/3..) enables a natural scaling procedure from which physical
properties of real quasicrystals may be deduced.

\vspace{10pt}

\hspace{\parindent}We will construct the basic idea of electronic
states near the Fermi energy with help of the knowledge of the Fermi
surface of crystalline approximants.

\vspace{10pt}

\hspace{\parindent}Anomalous transport
 in quasicrystals stems from non ballistic quantum diffusion
between electronic eigenstates, and has been shown to
 prevail even at higher
temperatures. From scaling analysis of the Kubo formula of electronic
conductivity, power law exponents of temperature dependence
conductivity will be found qualitatively. A discussion on
 metal-insulator in quasicrystals will conclude the paper.

\vspace{10pt}

\section{ Electronic structure and
Fermi surfaces of quasicrystalline approximants}

\vspace{10pt}

\hspace{\parindent}Quasicrystals are highly ordered
and anomalously resistive when compared to their amorphous
counterparts. Due to a lack of periodicity, an analytic
form of the Fermi surface ({\cal FS}) cannot be written down. On the other
hand,
it is now well established that some quasicrystalline
approximants share similar anomalous properties with
 pure icosahedral or decagonal phases, so that strict quasiperiodicity may
not be an unavoidable requirement to account for such physical properties
.~\cite{Approximants-I} It is therefore interesting to investigate
the Fermi
surfaces of approximant phases, which are well-defined and easier to calculate.

\vspace{10pt}

\hspace{\parindent}Fermi surface in metallic alloys is indeed, a meaningful
concept, enabling the determination of most of the physical properties which
depend, in some way, on the behaviour of conduction electrons. It turns
out that magnetic field can be used as a powerful probe of the shape
 of the Fermi surface.~\cite{Schoenb} For instance,
de Haas-van Alphen -dHvA- (magnetization) and Shubnikov
(magnetoresistivity) low temperature experiments yield to reliable and
quantitative geometrical informations such as radii or cross-sectional
areas, whereas magnetoresistance at high fields, reveals
whether or not open orbits are allowed in the plane normal to the magnetic
field. Nevertheless, the role of open electronic
orbits is quite complicated to figure out as they are thought to entail
additional quantum interference effects.~\cite{Stark-I}

\vspace{10pt}

\hspace{\parindent}DHvA shows up as
1/H-periodic oscillations in the magnetization, but is
usually observable in crystals of high structural quality,
at low temperatures, and in relatively high magnetic fields (above 10
Tesla). From a semiclassical treatment of electron dynamics in a
uniform magnetic field, the oscillation frequency
can be related to the extremal electronic closed
orbit in the related {\cal FS}. If $S_{e}(E_{F})$ is an extremal
cross-sectional area of the {\cal FS} in a normal plane to the
magnetic field, then the frequencies ${\cal F}$ of dHvA oscillations is
 ${\cal F}=\hbar c/2\pi e S_{e}(E_{F})$ (in Tesla units), whereas
effective masses will be given by $m^{*}\sim \hbar^{2}
\frac{\partial}{\partial E} S_{e}(E_{F})$.

\vspace{10pt}

\hspace{\parindent}Some attempts have been made to probe
Fermi surfaces in pure icosahedral AlPdMn compounds~\cite{Evert-I} with
resistivity of $1250\mu\Omega$cm at room temperature. DHvA signature was
found but still needs to be confirmed. It has to be stressed that
the measured compounds are anomalously resistive, which is an
unfavorable circumstance for the observation of {\cal FS} by means of dHvA
methods.

\vspace{10pt}

\hspace{\parindent}In the following, first numerical
calculations of Fermi surfaces in several crystalline
approximants of the icosahedral
AlCuFe, AlCuFe-Si and AlMgZn are presented. We believe that
experimental investigation of the shape of {\cal FS} in less resistive
quasicrystalline approximants and comparison with theoretical
calculations, may facilitate the analysis for pure quasicrystals.

\vspace{10pt}

\subsection{Electronic bandstructure calculations}

\vspace{10pt}

\hspace{\parindent}Before presenting our results, let us recall the
basic features of electronic properties in approximants and
quasicrystals. The band structure of quasicrystalline approximants
 have been calculated by means of {\cal LMTO} self consistent
method.~\cite{Fujiwara-II,Fujiwara-III} A pseudogap
with a width $\sim$ 0.5 eV existing at the Fermi energy,
 attributed to Hume-Rothery distortion, is supposed to be responsible
for the stability of the quasicrystals.~\cite{Fujiwara-III,Belin-I} This
pseudo-gap
 has been found theoretically in models of approximants, and experimentally
in approximants or pure icosahedral quasicrystals, so that
electronic properties are comparable.

\vspace{10pt}

\hspace{\parindent}It has also been argued that the
 pseudogap could be widened and deepened by aggregation of icosahedral
clusters or their hierarchical aggregation.~\cite{Trambly-I,Janot-II}
Concerning the total electronic density of states (TDOS), self-consistent
ab-initio calculations describe TDOS as a dense set of very sharp spikes
with a width less or equal to 0.01 eV~\cite{Fujiwara-II,Fujiwara-III},
sensitively depending upon the scattering length.
 The spikes may be due to quasiperiodicity, cluster aggregation and
{\it d}-orbital resonance.~\cite{Fujiwara-II,Fujiwara-III,Trambly-I}
Nevertheless, due to finite experimental resolution and averaging information over
wide spatial area, these features are difficult to observe
experimentally. Very Recent low temperature tunneling experiments 
in AlNiCo decagonal quasicrystals seem to be consistent with such
spiky peaks \cite{Guohong}.

\vspace{10pt}

\hspace{\parindent}Concerning transport properties, it is expected that
all the quantities, involving a surface integral
 over the whole {\cal FS}, will be affected substantially,
if anisotropic effects generate increasing distortion from ideal
spherical {\cal FS}. An illustration has been given in
 $\hbox{Al}_{62.5}\hbox{Cu}_{25}\hbox{Fe}_{12.5}$-FCI quasi-crystals, by S.E. Burkov et
al.,~\cite{Burkov-I} who proposed that limited number of Bragg peaks
(giving rise to the strongest pseudopotential)
might intervene in the Fermi surface-Brillouin zone interaction mechanism.
Taking into account the contribution of 42 Bragg planes, an alteration of
the shape of spherical Fermi surface was speculated, and a corresponding
reduction of electronic conductivity, in the nearly free electron
framework, estimated. This heuristic approach reveals how
icosahedral symmetry contribute in reducing conductivity from the
isotropic case.

\vspace{20pt}

\subsection{Models and Results}

\subsubsection{$\omega$-{\rm AlCuFe}}

\vspace{20pt}

\hspace{\parindent}Results from {\cal LMTO} band structure calculations for
several models are presented. The first one is the $\omega$-phase of a
crystal AlCuFe (tetragonal symmetry) 
whose electronic properties are metallic. The electronic
structure of this crystal has already been computed, by ab-initio
{\cal LMTO} methods.\cite{Trambly-II}

	
\vspace{20pt}

\hspace{\parindent}On Fig. 1. the total density of states (TDOS) and the Fermi
surface in the $K_{z}=0$ plane for the $\omega$-phase of AlCuFe are
shown ( $K_{x},K_{y}$ are given in $2\pi/a$ units,
with $a=6.336\AA$). The Fermi surface shows the symmetries of AlCuFe alloy and is
compatible with band dispersion. One notes that this Fermi surface
seems to have only closed orbits, and for instance, five shells of closed
electronic orbits around the $\Gamma=(0,0)$ point. To estimate the
associated dHvA frequencies, one has to calculate
$\oint_{\gamma_{s}}K_{y}dK_{x}$ ($\gamma_{s}$ the countour of the
closed orbits). A rough estimation
of dHvA frequencies of two quasi-circle shaped orbits gives
${\cal F}\sim 15, 79$ Tesla. The principal frequencies may appear in
oscillations of magnetization M(1/B). One notes however there are many
``breakdown points'' which can couple, because of tunneling, these different closed regions of
the FS.

\vspace{10pt}

\hspace{\parindent}The results obtained for this crystal are important 
since they allow a direct comparison of electronic structure between
metallic and anomalous transport compounds. However, one
stresses that it may not be sufficient to compare respective
bandstructure since electronic transport also requires knowledge of
dynamical information that may be only indirectly related to static
spectral properties.


\vspace{20pt}

\subsubsection{{\rm 1/1 AlCuFe}}

\vspace{20pt}

\hspace{\parindent}The bandstructure of 1/1 approximant of i-AlCuFe,
based on a hypothetical structural model \cite{Cok}, has been
evaluated and is in good agreement with previous calculations.
\cite{Trambly-II} On Fig. 2., the section of the {\cal FS} in the
plane $K_{z}=0$ and the TDoS are presented. We estimate four dHvA
frequencies, associated to closed orbits, which are respectively 
1.44, 4, 7.9, and 13 Tesla ($K_{x},K_{y}$ are given in $2\pi/a$ units
with $a=12.30\AA$).


\vspace{10pt}

\hspace{\parindent}Bandstructure was computed and nearly dispersionless
 bands were obtained. Fermi velocity was estimated
$v_{F}\sim 0.5\times 10^{7} cm/s$ (several order of magnitude lower than the
corresponding values in amorphous systems), resulting in a mean free
 path much shorter than the interatomic distance
\cite{Fujiwara-III}. Accordingly, usual semi-classical treatment of
 transport may be unreliable. We will discuss later the relation
between bandstructure features and anomalous transport.

\vspace{20pt}

\subsubsection{{\rm 1/1 AlCuFe-Si}}

\vspace{20pt}

\hspace{\parindent}The lowest quasicrystalline
 approximant 1/1 of AlCuFe-Si (cubic phase with $a=12.336 \AA$) is
also investigated. The resistivity of this small approximant
as a function of temperature has been shown to be anomalous
 [$d\sigma(T)/dT>0$ (A. Quivy et al.~\cite{Berger-II})],
similar to that of the icosahedral phase of AlCuFe.
The model, we  used, has been taken from Mizutani et
al.~\cite{Mizutani-I}. By means of Rietveld method
for the powdered X-ray diffraction spectrum, the
approximant was found to form a cubic lattice ($a=12.336\ \AA$) with
the space group $Pm\bar{3}$. Two pseudo-Mackay icosahedral atomic clusters
$Cl_{\ 1}$ (with vacancy in the center) and $Cl_{\ 2}$ (with Cu atom in
the center) were identified, with 139 atoms in the unit cell.
Nonetheless, out of this 139 atoms, about 36 position sites were
 given with occupancy of $50\%$ by Fe and $50\%$ by Cu.
To circumvent such uncertainties, we have investigated different
 models and found that the only proper one, according to the bandstructure
results and
concentration stoichiometry, is given by considering that $Cl_{\ 1}$
consists of Cu with 1.0 occupancy in the $M_{\ 8}$ subcluster, and
Fe with 1.0, in the $Mn_{\ 1}$ subcluster (see reference \cite{Mizutani-I}
for details), whereas the situation in $Cl_{2}$ is reversed.


\vspace{20pt}

\noindent
Results for the 1/1 approximant of
$\hbox{Al}_{\ 55}\hbox{Cu}_{\ 25.5}\hbox{Fe}_{\ 12.5}-\hbox{Si}_{\ 7}$ as described
in the structural model proposed by Mizutani \cite{Mizutani-I} are presented
on Fig. 3. The TDOS is small at
 Fermi level (pseudo-gap), and the typical spiky
structure is also found. Corresponding electronic bandstructure
 in the vicinity of the Fermi level exhibits a high density of flat bands. 
The associated Fermi surface (in $K_{z}=0$
plane) is rather different from the one of the tetragonal phase of AlCuFe,
 in particular by the presence of open orbits. Smallest
closed areas (around $\{K_{x}=0,K_{y}=0\}$) correspond to frequency of about 
${\cal F}\sim 0.7 $ Tesla, whereas
closed orbits centered around $\{ K_{x},K_{y}\}=\{0.5,-0.5\}$ correspond 
to ${\cal F}\sim 6.38 $ Tesla ($K_{x},K_{y}$ are given in $2\pi/a$ units
with $a=12.336\AA$).


\vspace{20pt}

\subsubsection{{\rm 1/1-AlMgZn}}

\vspace{20pt}

\hspace{\parindent}Finally, electronic bandstructure and Fermi
surfaces of AlMgZn approximants (bcc, $Im\overline{3}$) have been
computed by varying atomic compositions in model derived from
Rietvelt analysis.~\cite{Mizutani-I} The band structure of
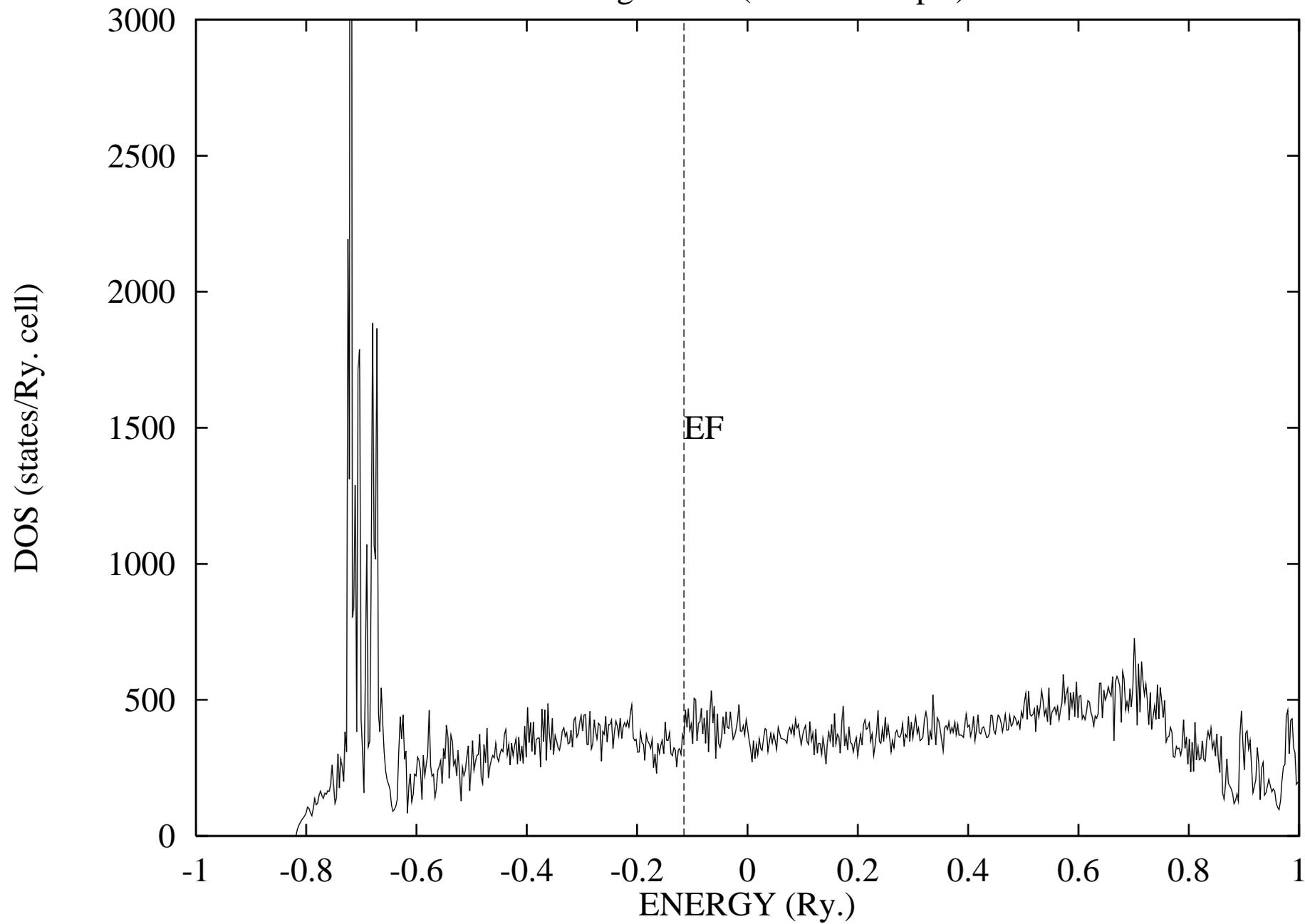
$\hbox{Al}_{45}\hbox{Mg}_{40}\hbox{Zn}_{15}$ 
near the Fermi level is shown on Fig. 4. Eight
bands cross the Fermi energy and each band has large energy dispersion
between $\Gamma=\{0,0\}$ and $H=\{K_{x}=0,K_{y}=-0.5\}$ points
 (i.e. the (100) direction), which enable to determine whether the
orbits of Fermi surfaces are related to hole or electron pockets.

\vspace{20pt}

\hspace{\parindent}One notes
that the $x,y,z$ axes are equivalent due to the $120^{\circ}$
rotation, but there is no $\pi/2$ rotation symmetry along x-,y- or z-
axes. The large energy dispersion is due
to contact of atom clusters along the (100), (010), (001)
directions.\cite{Fujiwara-III} Furthermore, they are many small
 electron and hole pockets over the whole region.


\vspace{20pt}

\hspace{\parindent}On Fig. 5, the corresponding TDoS and FS in the
$K_{x}=K_{y}$  plane are also given. We precise that 
$-1\leq K_{x},K_{y}\leq 1$ are given in units $2\pi/a, a=14.498\AA$.
Six bands provide several closed orbits whereas one
band forms an open orbit along the (100) direction. Very small
electron and hole pockets, may be rather difficult to observe given
the damping effects due to finite electron scattering
time. Notwithstanding we estimate the dHvA frequency and the effective 
mass for the largest orbit surrounding the $\Gamma$ point.  From the
Fermi surface we estimate ${\cal F}\sim 11 $ Tesla and by
considering bandstructure along the $\Gamma - $H direction
(H=$\{K_{x}=0,K_{y}=-1\}$), we find that the
corresponding effective mass is of order $m^{*}\sim 5m_{e}$ ($m_{e}$
the free electron mass). 


\vspace{20pt}

\subsubsection{Discussion about Fermi surface and relation to
anomalous transport}

\vspace{20pt}

\hspace{\parindent}Our studies have shown
that many hole and electron pockets emerge in the energy constant
contours of the Fermi surfaces. A noticeable feature for the FS of the
quasicrystalline approximants seems to be the existence of open orbits which
may be manifested through magnetic breakdown at high magnetic field. 
Methods such as dHvA at low temperature or angle-resolved photoemission spectroscopy, 
may be suitable to investigate such disconnected Fermi surfaces. 

\vspace{20pt}

\hspace{\parindent}One notes that the band dispersion of the
quasicrystalline approximants is so weak that $v_{F}=1/\hbar
\partial E_{n}(k)/\partial k$ turns out to be very small. Another noteworthy
feature allows to point out a difference between metallic
transport and anomalously reported quasicrystalline one.\cite{Roche-I} Elastic
scattering time $\tau_{el.}$ is generically related to an impurity
concentration in the system, enabling a semi-classical description of the dynamics of
wave packet in between scattering events. The exact details of
 bandstructure being usually neglected. In case of approximants, it
turns out that energy spacing between bands, in the vicinity of Fermi
energy, becomes very small, so that one has to define a typical
 time associated to $\hbar/(E_{n}-E_{m})$ (average between two
bands m and n) as a limit given by Heisenberg uncertainty principle.
If there exists a finite number of bands within
the energy window defined by a finite $\tau_{el.}$, then
transport may turn out to be anomalous because tunneling occurs
between different bands, which cause the instability of the wave
packet coherence. Actually, if one considers the crystalline phase
$\omega-\hbox{AlCuFe}$  and the 1/1 approximant of
icosahedral AlCuFe,AlCuFe-Si,AlMgZn, one
finds that, for $\tau_{el.}\sim 10^{-15}s$ (typical value found in
amorphous of similar resistivities), the energy window is $\Delta
E\sim 0.675 eV$ and confines about ten bands for the
approximants whereas the crystalline phase presents no bands. One notes
that the considered value for the scattering time may be 
questionable since it is usually deduced from experiment using
classical theories (which should be unapplicable in
case of anomalous transport). Notwithstanding, the aforementioned difference in the
bandstructure does not critically depends on the considered value of $\tau_{el.}$.

\vspace{20pt}

\hspace{\parindent}As real experiments do not provide 
quantitative informations about the bandstructure, a criterion for
occurrence of anomalous transport may be defined from detailed analysis of the
topology of the Fermi surfaces, by comparison between theoretical
results and experiments. For instance, high density of flat bands, in the vicinity
of Fermi level may promote the occurrence of open orbits. If such
a statement is valid, magnetic breakdown will be observed more
frequently for quasicrystalline approximants.


\vspace{10pt}

\section{Anomalous transport properties}

\vspace{10pt}

\subsection{ Anomalous diffusion and
temperature dependence of the conductivity}

\vspace{10pt}

\hspace{\parindent}Unusual electronic transport in quasicrystals is
 supposed to originate from anomalous quantum diffusion and
specific local atomic order. To circumvent the
complexity of the structural issues, when considering electronic
localization, one may address separately the effects of quasiperiodic
potential on electron dynamics, and
bandstructure effects, generic of approximants. In principle,
scaling analysis based on crystalline approximants should allow a
connection between both approaches.

\vspace{10pt}

\hspace{\parindent}In ordinary metals, neglecting weak localization
effects, one usually distinguishes two
different regimes for temperature dependent transport (even if the
frontier, given by Debye temperature,
 is by no means universal, but rather alloy-dependent). To
calculate $\sigma(T)$, one must treat properly the electron-phonon
interaction within the usual linear response scheme. The Kubo formula is
a suitable starting point, and can be written as:~\cite{Kubo}

\vspace{10pt}

$$\sigma(T)=\frac{e^{2}n_{0}}{m}\int_{-\infty}^{+\infty}\biggl
( \frac{\partial f(E)}{\partial E}\biggr)\times
\frac{\Gamma(k_{F},E\pm i\eta)}{2\Im m \Sigma(k_{F},E)}dE$$

\vspace{10pt}

\noindent
with $n_{0}$ the electronic density,
 $f(E)=(1+e^{\beta(E-\mu)})^{-1}$ the Fermi-Dirac function and
$\Gamma(k_{F},E\pm i\eta)$ a so-called vertex function
which accounts for all the
collision process of the {\it propagating electrons} with the
phonons. $\Sigma(k_{F},E)$ is related to the transition rate of one
electron in one eigenstate, due to the coupling with the thermal
perturbation.

\vspace{10pt}

\hspace{\parindent}The evaluation of the Kubo formula is done by summing the
relevant ladder diagrams for electron-phonon interactions,
 and reducing the vertex function to an integral equation, that is solved numerically.
~\cite{Mahan} In the case of weakly disordered metals, the vertex
function reduces to 1, whereas inelastic relaxation time
$\frac{1}{\tau_{i}(E)}=-2(\Im m \Sigma(k_{F},E))^{-1}$ is given by
Fermi's golden rule. It is shown that
in the temperature regime such that $\hbar \omega_{D}\gg k_{B}T$
(with $ \omega_{D}\sim c_{S}\frac{\pi}{a}$, $c_{S}$ the sound velocity
and $a$ the lattice constant), phonons with energies $\hbar\omega\sim
k_{B}T$ will play a dominant role. Transport time 
will be given accordingly by \cite{Abrikosov}

$${\displaystyle \frac{1}{\tau_{tr}}\sim \frac{T}{\hbar}
\biggl(\frac{T}{\hbar\omega_{D}}\biggr)^{4}\sim T^{5}}$$

\noindent
and from Drude formula ${\displaystyle
\frac{e^{2}n_{0}}{m}\tau_{tr}}$,
Bloch-Gruneisen law will follow~\cite{Abrikosov}:

$${\displaystyle \sigma_{DC}(T)\sim
\frac{n_{0}e^{2}\tau_{tr}}{m}\sim T^{-5}}$$

\vspace{10pt}

\hspace{\parindent}Electronic
transport properties in quasicrystals or approximants
turn out to be {\bf non-metallic}, opposite Mathiessen rule being one of the
most spectacular manifestation.~\cite{Mayou-I,Berger-I,Honda-I} Experiments
reveal that the general form of conductivity is generically $\sigma_{DC}(T)\sim \sigma_{4K}+\Delta T$
with $\Delta \sigma(T)\sim T^{\alpha} \ \
\alpha\in[1,1.5]$.~\cite{Gignoux-I,Kimura-I,Tamura-I,Lalla-I} Theoretically,
 recent studies of anomalous electronic transport in presence of
 inelastic collisions, by Schulz-Baldes and Bellissard~\cite{Bellissard-I},
have strengthened the heuristic arguments for the suggested {\bf anomalous Drude
law} ($\sigma\sim \tau^{2\beta-1}$) \cite{Roche-II,Mayou-I}, where
$\tau$ is supposed to be the relevant inelastic scattering rate at a
given temperature. The anomalous Drude law has been shown to account
for all the specific electronic properties of quasicrystals.

\vspace{10pt}

\hspace{\parindent}In the following, temperature dependent power law is shown to be 
consistent with a scaling analysis of the Kubo formula, in the low temperature
regime. From the knowledge of exact
eigenstates $|\alpha\>$, in the vicinity of Fermi energy,
the Kubo formula is evaluated from $\sigma_{DC}= \int_{-\infty}^{+\infty}
 dE {\small  (-\partial f(E)/\partial E) }
\sum_{\alpha}\delta(E-E_{\alpha}){\cal D}_{\alpha}(E)$ where

$${\cal D}_{\alpha}(E)=-\frac{1}{\pi}\lim_{\gamma\to 0^{+}} \Im
m\<\alpha|\hat{v}_{x}\frac{1}{E-{\cal H}+i\gamma}\hat{v}_{x}|\alpha\>$$


\noindent
is the contribution to diffusivity of a given eigenstate $|\alpha\>$. The
cut-off $\gamma$ stands for the relevant relaxation time that will
dominate the transport process. Scaling analysis of the Kubo formula
has been performed for realistic models of approximants of the decagonal
AlCuCo quasicrystals, including sp-d orbitals of all the atomic
species. This leads to a self-consistent treatment of sp-d hybridization effects,
 which are known to produce important alteration of a naive nearly
free electronic transport model. Anomalous power
law dependence of diffusivity in $\gamma$ was found ~\cite{Fujiwara-III,Fujiwara-IV}

$${\displaystyle \langle {\cal
D}\rangle \sim \gamma^{\nu}\sim
\biggl(\frac{1}{\tau_{tr}}\biggr)^{\nu},\  \nu\simeq
0.25}$$

\noindent
where $\gamma$ has to be replaced by a suitable scattering rate
dominating transport mechanism. We stress that this result
gives the first ``quantitative exponent'' characterizing the
transport coefficients in the vicinity of the
 Fermi level for a realistic model. From Kubo's formula, the temperature dependent part of the conductivity is related, in a first
approximation, to the transport time. In the low temperature limit,
using the Debye model, one expects Bloch-Gruneisen law to be
applicable. Then from $\tau_{tr}\sim T^{\mu}$
with $\mu=-5$, we conclude that :

$${\displaystyle
 \Delta\sigma(T)\sim T^{\eta},\ \ \hbox{\small with}\ \  \eta\simeq 1.25}$$

\vspace{10pt}

\noindent
This is in good agreement with the values obtained
experimentally for several quasicrystalline and approximant
phases at temperature $\geq
400mK$. \cite{Gignoux-I,Kimura-I,Tamura-I,Gignoux-II} 
For instance, in icosahedral $\hbox{Al}_{70}\hbox{Pd}_{20}\hbox{Re}_{10}$ bulk
samples, $\hbox{Al}_{70}\hbox{Pd}_{18}\hbox{Re}_{12}$,
 $\hbox{Al}_{70}\hbox{Pd}_{22}\hbox{Re}_{8}$ ribbon samples
($R=\rho_{20K}/\rho_{300K}\simeq 2-10$), and
$\hbox{Al}_{65}\hbox{Cu}_{20+x}\hbox{Ru}_{15-x}$ ingots, the
transport exponents are respectively  $\eta=1.3$ and
$\eta=1.4,1.45$, \cite{Kimura-I,Tamura-I} and $\sim 1.3$
\cite{Lalla-I}, for temperature range $[400mK,300K]$. One
recalls that in quasicrystals, the typical Debye temperature, which
limits the ``low temperature regime'' estimated from
 specific heat measurements are typically $\propto 400\sim 500K$.~\cite{Martin-I}

\vspace{10pt}

\hspace{\parindent}One also notes that 
Kubo formula  $\sigma\sim \tau^{2\beta-1}$ with $\beta=0.375$ in the d-AlCuCo
approximants, is also consistent with opposite Mathiessen rule. 
If scattering time increases (improving structural order or
decreasing temperature), conductivity decreases. 

\vspace{10pt}

\hspace{\parindent}In summary, scaling analysis of the Kubo formula, combined to
temperature dependence of transport time, has been shown to be consistent
with the power-laws found in experiments. The main asset of scaling
analysis with 3D realistic models of periodic approximants is that
 complicated effects due to local atomic order (sp-d hybridization,
local cluster geometries) have been integrated self-consistently into the
ab-initio calculations. This point is critical, given that
 metal-insulator transition can be induced by a tiny variation of
atomic stoichiometry, while the material preserves long
range order. For instance, a metal insulator transition
 has been studied in the $\hbox{Al}_{70.5}\hbox{Pd}_{21}\hbox{Re}_{8.5-x}-\hbox{Mn}_{x}$ quasicrystalline
compounds.~\cite{Poon-MIT} The signature of such a transition 
occurs as concentration of manganese is increased from
$x=0$ to $x=5$, whereas long range quasiperiodic order remains
unchanged. Finally, from studies of electronic
properties in approximants, one may relate non-metallic transport character to
specific features of Fermi surface combined to scaling analysis of the 
transport coefficients.

\vspace{20pt}

\subsection{Discussion on low-temperature transport
and metal-insulator transition}

\vspace{20pt}

\hspace{\parindent}From experiments, one can basically distinguish two
low-temperature transport regimes for
quasicrystals. The anomalous power law 
temperature dependence of the conductivity appears as a criterion for
distinguishing both regimes.

\vspace{10pt}

\hspace{\parindent}Indeed, in some quasicrystals, low temperature
transport is dominated by quantum interferences effects (QIE), 
as described for weakly disordered systems, and leads to a finite conductivity
at zero temperature, and to typical temperature and field dependence
 of resistivity. Detailed studies have been carried out in
particular in pure quasicrystalline phases of AlCuFe, AlPdMn, AlCuCo,
AlNiCo.~\cite{Berger-I,Martin-I,Poon-I,Rapp-I} From data analysis and
theories of QIE, an estimation of mean free path about
$20\sim 30 \ \AA$ was extracted. However, the origin of coherence
 of wavefunctions in quasicrystals may be different from that given by the conventional theory of QIE, which relies on quantum interference
between propagating and backscattered waves.  Coherence in
quasicrystals may be rather due to the ``quasiperiodic mesoscopic order'' as recently shown by scaling analysis of electronic participation 
ratio in realistic structural models.~\cite{Fujiwara-IV}

\vspace{10pt}

\hspace{\parindent}Another transport regime at low temperatures
corresponds to stronger electronic localization induced by quasicrystalline
order, and yields to better insulating quasicrystals
 at zero temperature~\cite{Gignoux-I,Gignoux-II}.
Besides, for largest values of ${\cal R}=\rho_{4K}/\rho_{300K}\sim
120$, the associated power law ${\displaystyle
\Delta\sigma(T)\sim  T^{\eta}}$ seems to prevail down to very low temperature $T
\downarrow 4K$.~\cite{Gignoux-I} This bevahiour is reminiscent of
 scale invariance, as usually found in systems close to
metal-insulator transition. No weak localization effects are
revealed from experiments.

\vspace{10pt}

\hspace{\parindent}Furthermore, at $4K\leq T\leq
16K$, the conductivities of some samples of icosahedral AlPdRe have been
 successfully described by means of McMillan scaling theory, which treats
localization and electron-electron correlation on equal
footing.~\cite{McMillan} Experiments by Lin et al.~\cite{Lin-MIT}
 have been interpreted through $\sigma(T)=\sigma_{4K}+\sqrt{T/\Delta}$
where $\Delta$, the Coulomb gap (the relevant energy scale in the insulating phase, 
$\Delta\propto\rho_{4K}^{2}$), is determined by conductivity and electron
tunneling measurements. This suggests that, unlike
disordered amorphous, the metal insulator transition in quasicrystals
may be induced by the atomic order on local or mesoscopic scale. The
exact physical origin of such a metal-insulator transition
is still a challenging unresolved problem.

\vspace{10pt}

\hspace{\parindent}One may depict the transition from one
low-temperature transport regime to another in the following way. Quasicrystalline phases
that are dominated by QIE at low temperatures, may suffer screening by
tiny impurities and/or disruptions of the relevant mesoscopic
atomic order, responsible for stronger geometrical localization
(defined by ``resonances'' as described in \cite{Roche-I,Trambly-I}).
 An experimental evidence is provided by the deviations
observed in AlPdMn~\cite{Gignoux-I} and AlCuRu~\cite{Kimura-I}
at about 80K, where typical power law for conductivity can be
recovered.

\vspace{20pt}

\section{Conclusion}

\vspace{20pt}

\hspace{\parindent}Experiments reveal that quasicrystals present
remarkable electronic transport properties, hitherto unexpected. In particular,
 a metal-insulator transition is approached, by improving
 quasicrystalline order. Within the one electron scheme, several
 theoretical approaches suggest an intermediate type of algebraic
localization between purely extended states and exponentially
localized states.~\cite{Kohmoto-QP,Macia,Roche-IV,Roche-II}
Different schemes were proposed to explain original
 transport properties, amongst which, non-ballistic quantum diffusion
 for long relaxation times or hopping mechanisms induced by
 interband transitions at shorter ones. In this paper, we have shown how
scaling analysis of the Kubo formula may lead to power-law temperature
 dependence of electronic conductivity in the low temperature regime.

\vspace{20pt}

\hspace{\parindent}Features of the Fermi surfaces of several
quasicrystalline approximants were presented based on ab-initio
calculations. They may hopefully support further experimental
investigations of real {\cal FS} in quasicrystalline materials, and improve our
understanding about non-metallic transport in approximants and quasicrystals.

\vspace{20pt}

\section{ Acknowledgments}
 The authors would like to acknowledge G. Trambly de Laissardi\`ere
and T. Ziman for useful comments. One of us
(S.R.) is indebted to the European Commission and the Japanese
 Society for Promotion of Science (JSPS) for joint financial support
(Contrat ERIC17CT960010), and to the Department
of Applied Physics of the University of Tokyo for its kind hospitality.

\section{Figures captions}

\vspace{20pt}

\noindent
Fig. 1. Total density of states and Fermi surface in the Brillouin
zone ($K_{z}=0$ plane) of the $\omega$-AlCuFe crystalline phase

\vspace{20pt}

\noindent
Fig. 2.  Total density of states and Fermi surface 
 of the 1/1-AlCuFe quasicrystalline phase.

\vspace{20pt}

\noindent
Fig. 3. Total density of states and Fermi surface of 1/1-AlCuFeSi.

\vspace{20pt}

\vspace{20pt}

\noindent
Fig. 4.  Electronic bandstructure of 1/1-AlMgZn quasicrystalline approximant.

\vspace{20pt}

\noindent
Fig. 5. TDoS and Fermi surface section of 1/1-AlMgZn in the plane
$K_{x}=K_{y}$.

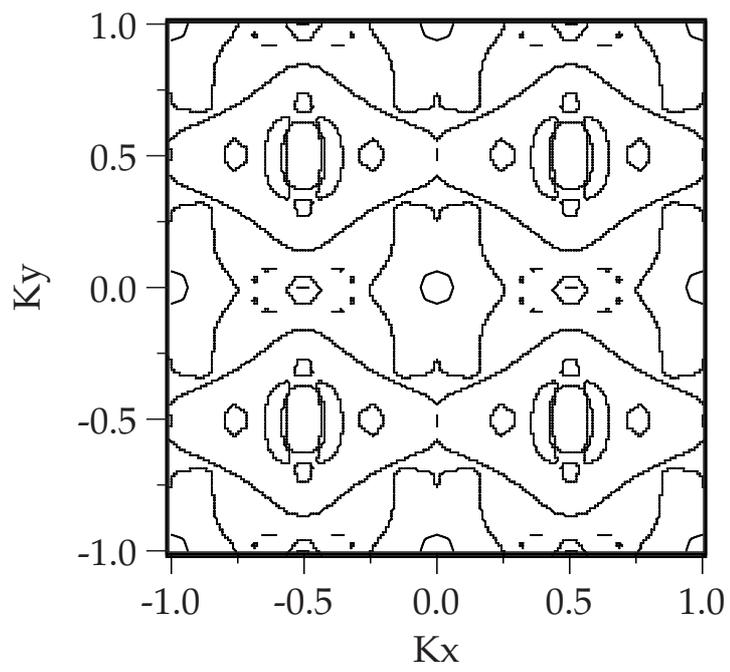
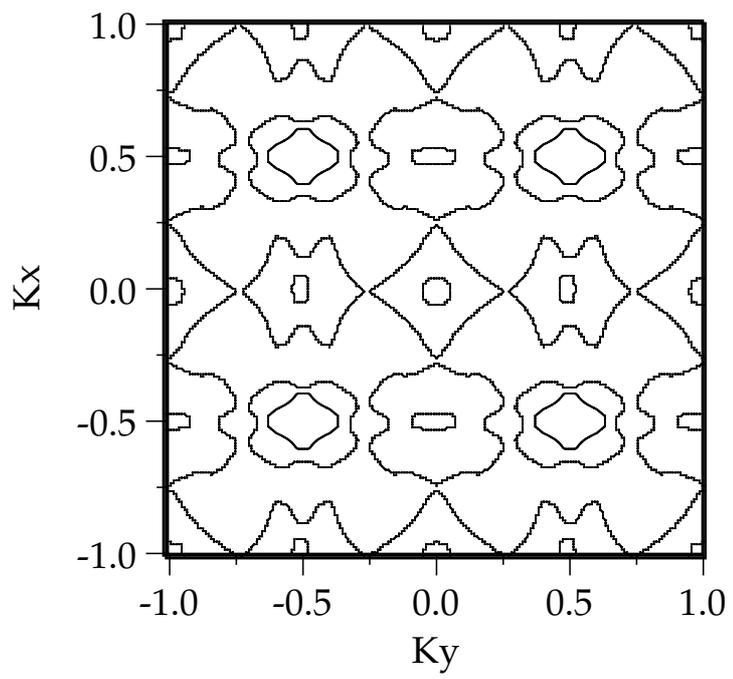
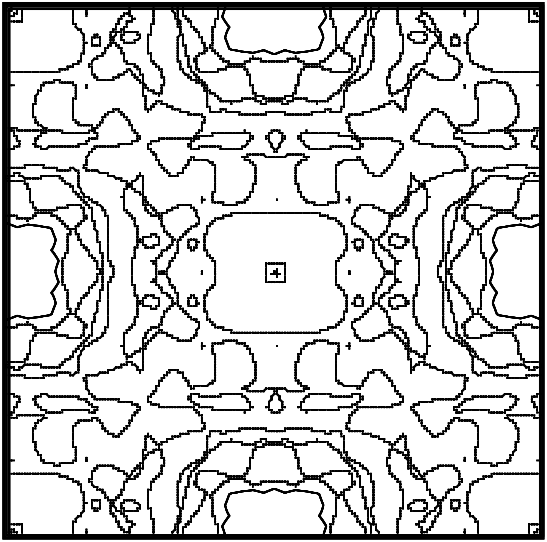
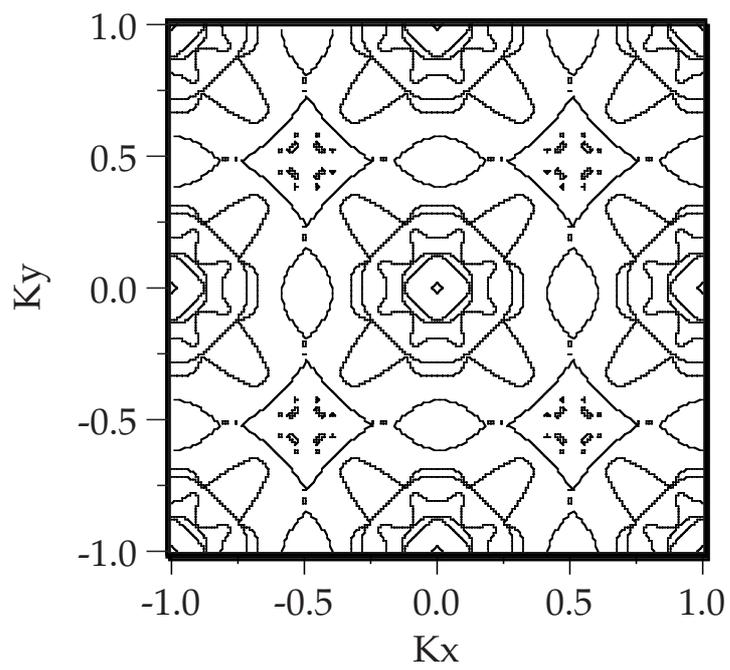
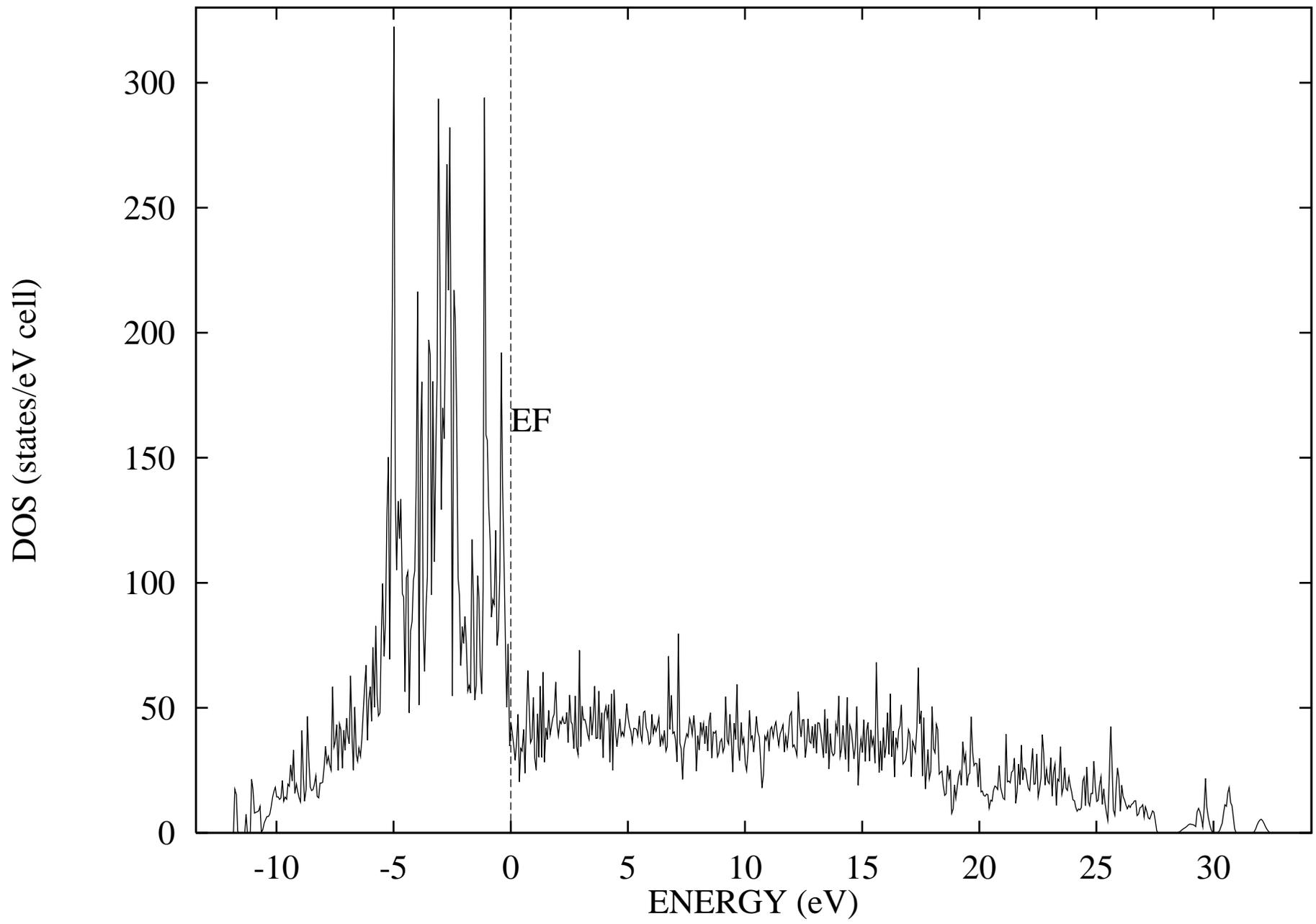
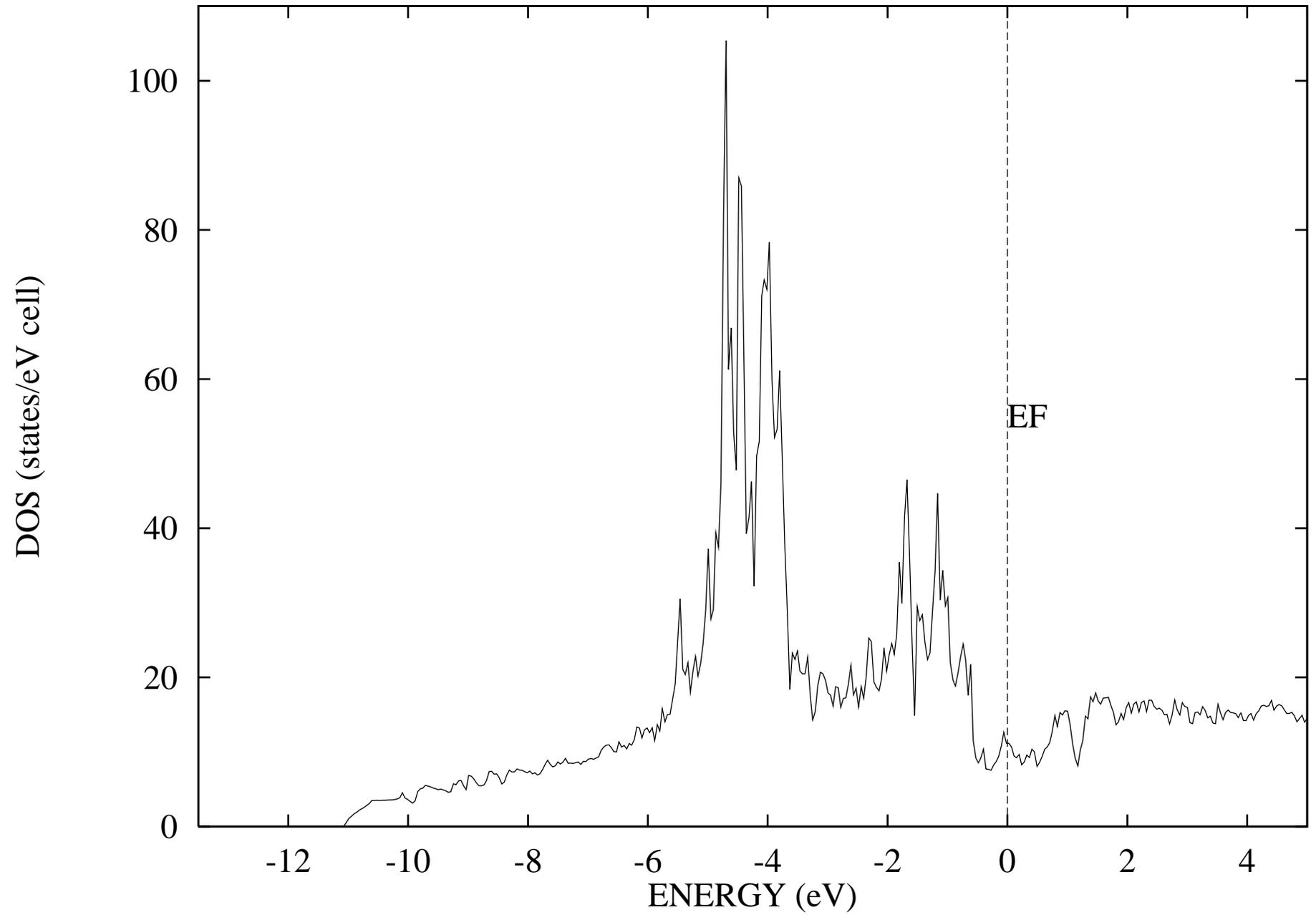
\end{document}